\pdfoutput=1

\documentclass[%
reprint,
superscriptaddress,
nofootinbib,
amsmath,amssymb,
aps, pra,
longbibliography
]{revtex4-1}

\usepackage[colorlinks=true,urlcolor=blue,citecolor=blue,linkcolor=blue]{hyperref}
\def\ket#1{| #1 \rangle}
\def\bra#1{\langle #1 |}
\def\bracket#1#2{\langle #1 | #2 \rangle}
\def\kb#1#2{| #1 \rangle\!\langle #2 |}

\def\cN{\mathcal{N}}
\def\cO{\mathcal{O}}

\def\eq#1{Eq.~\eqref{eq:#1}}

\def\eq#1{Eq.~\eqref{eq:#1}}

\def\ket#1{| #1 \rangle}
\def\bra#1{\langle #1 |}
\def\bracket#1#2{\langle #1 | #2 \rangle}
\def\kb#1#2{| #1 \rangle\!\langle #2 |}

\def\eq#1{Eq.~\eqref{eq:#1}}

\begin{document}

	\title{Quantum Algorithm for Spectral Measurement with Lower Gate Count}
	
	\author{David Poulin}
	\affiliation{D\'epartement de Physique \& Institut Quantique, Universit\'e de Sherbrooke,  J1K 2R1, Canada}%
	\affiliation{Canadian Institute for Advanced Research, Toronto, ON M5G 1Z8, Canada}
	\author{Alexei Kitaev}
		\affiliation{California Institute of Technology, Pasadena, CA 91125, USA}
	\author{Damian S. Steiger}
	\affiliation{Theoretische Physik, ETH Zurich, 8093 Zurich, Switzerland}%
	\author{Matthew B. Hastings}
	\affiliation{Station Q, Microsoft Research, Santa Barbara, CA 93106-6105, USA}
	\affiliation{Quantum Architecture and Computation Group, Microsoft Research, Redmond, WA 98052, USA}
	\author{Matthias Troyer}
	\affiliation{Theoretische Physik, ETH Zurich, 8093 Zurich, Switzerland}%
	\affiliation{Quantum Architecture and Computation Group, Microsoft Research, Redmond, WA 98052, USA}

	\date{\today}
	
	\begin{abstract}
		We present two techniques that can greatly reduce the number of gates required to realize an energy measurement, with application to ground state preparation in quantum simulations. The first technique realizes that to prepare the ground state of some Hamiltonian, it is not necessary to implement the time-evolution operator: any unitary operator which is a function of the Hamiltonian will do. We propose one such unitary operator which can be implemented exactly, circumventing any Taylor or Trotter approximation errors. The second technique is tailored to lattice models, and is targeted at reducing the use of generic single-qubit rotations, which are very expensive to produce by standard fault tolerant techniques. In particular, the number of generic single-qubit rotations used by our method scales with the number of parameters in the Hamiltonian, which contrasts with a growth proportional to the lattice size required by other techniques.
	\end{abstract}
	
	\maketitle
	
	The simulation of quantum systems is one of the main foreseen  applications of quantum computers \cite{feynman1982simulating, lloyd1996universal}, it can be used to predict the properties of materials \cite{abrams1997simulation,gubernatis,bauer2016hybrid,wecker2015solving}, molecules \cite{aspuru2005simulated}, quantum fields \cite{jordan2012quantum}, etc. There are two conceptual ingredients to a quantum simulation, corresponding roughly to statics and dynamics. The static part consists in preparing an initial state of physical interest. Because we are typically interested in low-energy properties, we will often require to prepare the quantum computer in the ground state $\ket{\phi_0}$ of the simulated system's Hamiltonian $H$. Dynamics consists in reproducing the effect of the time-evolution operator $U(t) = e^{-iHt}$. While methods to produce the time-evolution operator have been known for decades \cite{feynman1982simulating, lloyd1996universal}, the static problem is often a bottleneck of quantum simulations. 
	
	Simulating the dynamics of a quantum system on a quantum computer requires a quantum circuit which (approximately) reproduces the time-evolution operator $U(t) = e^{-iHt}$. The two standard approaches are the Trotter-Suzuki decomposition \cite{lloyd1996universal,jordan2012quantum} and Taylor  expansions \cite{berry2015simulating}, in addition to recent advances based on quantum signal processing \cite{low2017optimal}, all of which realize an approximation of $U(t)$ with controllable systematic errors. While the ideas presented in this Letter are primarily intended to simplify the static problem, the unary encoding technique we present can also be used to reduce the number of gates required to simulate time evolution.
	
	The ability to reproduce the dynamics $U(t)$ of a quantum system on a quantum computer also provides a means to solve the static problem. This is because, employing quantum phase estimation \cite{kitaev1995quantum, cleve1998quantum, abrams1999quantum}, the circuit implementing $U(t)$ can be used to perform a projective energy measurement.  Thus, given a good approximation $\ket{\tilde\phi_0}$ of the ground state $\ket{\phi_0}$,  an energy measurement should collapse the wavefunction onto the ground state with reasonable probability $|\bracket{\phi_0}{\tilde\phi_0}|^2$. 
	
The state $\ket{\tilde\phi_0}$ could be obtained from various approximations such as mean-field. When no simple good approximation $\ket{\tilde \phi_0}$ exists, we can express the Hamiltonian $H = H_0+V$ as a sum of some simple term $H_0$ of which we can easily prepare the ground state, and an ``interaction'' term $V$. Then, the Hamiltonian $H(g) = H_0 + g V$ has a simple ground state at $g = 0$ and its ground state at $g=1$ is the one we want to prepare. We can thus initialize the quantum computer in the ground state $\ket{\phi_0(g=0)}$ of $H(0)$, and perform a sequence of measurements which, with high probability, will result in the state $\ket{\phi_0(1)}$. The main idea \cite{somma2008quantum} is that for $|g-g'|$ sufficiently small, the probability $|\bracket{\phi_0(g)}{\phi_0(g')}|^2$ that a measurement of $H(g')$ performed on $\ket{\phi_0(g)}$ results in the ground energy should be close to 1. We thus choose a sequence of interaction strengths $0=g_0<g_1<g_2<\ldots <g_L = 1$  and perform a sequence of energy measurement using phase estimation of the time-evolution operator $U(j,t) = e^{-iH(g_j)t}$. As predicted by the quantum Zeno effect, a sequence of almost identical measurements should drag the state in the measurement basis with high probability $\prod_{j=1}^{L}|\bracket{\phi_0(g_{j-1})}{\phi_0(g_j)}|^2$. This method is equivalent to adiabatic evolution \cite{FGGS00a,AT03b}, but has the advantage that only time evolution under a {\em time-independent} Hamiltonian is required.
Other methods to prepare the ground state include quantum Metropolis sampling \cite{temme2011quantum, yung2012quantum}, and Grover search \cite{poulin2009preparing}.
	
In all of these approaches, the time-evolution operator is used jointly with quantum phase estimation to realize an energy measurement. This simply builds on the fact that $H$ and $U(t)$ share the same eigenstates $\ket{\phi_k}$ and, for sufficiently small $t \leq \pi/\|H\|$, have a one-to-one mapping between their eigenvalues, $E_k$ and $e^{-iE_kt}$ respectively. Clearly, this is true for many other function of the Hamiltonian $f(H)$, in particular functions with a Lipschitz continuous inverse at $E_0$.  The first technique we present uses ideas from quantum walks \cite{szegedy2004quantum}, Taylor  expansions \cite{berry2015simulating}, and the related Qubitization \cite{low2016qubitization} to build a unitary transformation $W$ which is a simple function of $H$, basically $W=\exp\{i \arccos(Ht)\}$. The unitary circuit $W$ does not represent any sort of physically meaningful evolution or transformation, but has the advantage of being implementable exactly in an ideal quantum circuit using roughly the same number of gates as required by a single Trotter step or a single segment of a first-order Taylor approximation. A key message of this Letter is that in order to get a good simulation of nature, it is not necessary to imitate nature: a quantum computer can sometimes employ physically inaccessible routes.

	The second technique we present  becomes motivated once we realize that not all gates in a quantum circuit have the same cost. In standard approaches to fault-tolerant quantum computation \cite{aharonov1997fault, shor1996fault, steane1999efficient, raussendorf2007fault}, Clifford gates (CNOT, Hadamard, $S =$ diag($1,i$)) are realized in an intrinsically fault-tolerant way. An extra gate is obtained by magic state distillation \cite{knill2004fault, bravyi2005universal, jones2013multilevel, haah2017magic, hastings2017distillation}. It can be either a Toffoli = control-control-not, or $T = $ diag($1,e^{i\pi/4}$), or $\sqrt{\rm SWAP}$, or control-SWAP. These gates share the property of belonging to the third-level of the Clifford hierarchy \cite{gottesman1999demonstrating}, so they can be distilled by standard methods and together with the Clifford gates they form a universal gate set. Thus, any single-qubit rotation can be approximated by a sequence of universal gates in a protocol called gate synthesis \cite{dawson2005solovay,KMM12a,ross2014optimal, bocharov2015prl, bocharov2015efficient}. The overhead associated with distillation and synthesis scales poly-logarithmically with the inverse targeted accuracy $\delta$, and the constant pre-factors are large, making the cost of third-level gates substantially larger than Clifford gates, and the cost of generic single-qubit rotation even more so.
	
	All the simulations methods mentioned above make use of the fact that the Hamiltonian is sparse in some natural basis, or typically that it can be written as a sum of terms $H = \sum_{j=1}^N \alpha_j P_j$ where $N$ grows polynomially with the system size $n$ (the number of particles or lattice size). For instance, if there can be at most $k$ qubits involved in each interaction term, $N$ would scale like $n^k 3^k$.  Since there are $N$ real parameters $\alpha_j$ in this Hamiltonian, the quantum circuit for $U(t)$ requires a number of generic single-qubit rotations that scales like $N$.   In lattice models however, such as encountered in condensed-matter physics or lattice field and gauge theories, many of the $\alpha_j$ have the same value.  For instance the Hubbard model $H_{\rm Hubbard} = -t \sum_{\langle i,j\rangle,\sigma} c_{i,\sigma}^\dagger c_{j,\sigma} + U \sum_j c_{i\uparrow}^\dagger c_{i\uparrow} c_{i\downarrow}^\dagger c_{i\downarrow}$ or the Ising model $H_{\rm Ising} = g \sum_j X_j + J\sum_{\langle i,j\rangle} Z_iZ_j$ each contain only a single real parameter $U/t$ and $J/g$ respectively. 
	
	The above parameter counting argument suggest that it may in principle be possible to realize the time-evolution operator of such systems using only a constant number of generic single-qubit rotations. The second technique we present realizes this expectation by replacing these $N$ generic single-qubit rotations by third-level gates, hence avoiding the cost of synthesis. If we denote by $C_D(\delta)$ the cost of distilling a third-level gate and $C_S(\delta)$ the cost of synthesising a generic single-qubit rotation, then the cost of our method is $KC_D(\delta)C_S(\delta) + N C_D(\delta)$ where $K$ is the number of $\alpha_j$ with distinct values, while each Trotter step costs $NC_D(\delta)C_C(\delta)$. Our method can straightforwardly be incorporated into the Taylor series method to obtain similar savings.
	
	\medskip
	\noindent{\em Spectrum by Quantum Walk} --- 
	To prepare the ground state of Hamiltonian $H$, our approach is to realize a simple quantum circuit $W$ which does not implement a unitary time evolution $U(t)$ but some other function of the Hamiltonian. Without loss of generality, we assume that the Hamiltonian is non-negative $H\geq 0$ and that it can be expressed as $H = \sum_{j=0}^N \alpha_j P_j$ where the $P_j$ are multi-qubit Pauli operators and $P_0 = I$. We rescale the Hamiltonian by a factor $\cN = \sum_{j=0}^N |\alpha_j| \in \cO(N)$ and note 
	\begin{equation}
	\bar H = \frac H{\cN} = \sum_j|\beta_j|^2 P_j,
	\label{eq:H}
	\end{equation}
	where $\beta_j = \sqrt{|\alpha_j|/\cN}$ and it follows that  $\sum_j |\beta_j|^2=1$. Note that any sign of $\alpha_j$ can be absorbed in the definition of $P_j$.  Obviously this rescaling does not affect the eigenstates, but it does change the spectral gap by a factor $\cN$ and this will be important when comparing this algorithms to ones based on Trotter expansions. 
	
	The unitary transformation $W$ we construct acts on  $n+\log (N+1)$ qubits, i.e., the $n$ {\em system} qubits and $\log (N+1)$ {\em control} qubits whose basis states $\ket j$ are in one-to-one correspondence with the $(N+1)$ terms $P_j$ of the Hamiltonian. There exists an invariant subspace of $W$ on which the spectrum of $W$ is a simple function of $H$. By initializing the quantum computer to that subspace, we thus obtain the desired effect.
	
Following \cite{berry2015simulating, low2016qubitization}, define $\ket \beta$, $B$, $S$ and $V$ as follows 
	\begin{align}
	\ket\beta &= B\ket 0 = \sum_j \beta_j \ket j, \\
	S &=B (I-2\kb 00)B^\dagger=(I-2\kb \beta\beta)\otimes I, \quad{\rm and}\\
	V&= \sum_j \kb jj \otimes P_j.
	\end{align}
The identities $S^2=V^2=I$ tell us that $S$ and $V$ are reflexions so they can simultaneously be put in block-diagonal form with blocks of size 2.   
Indeed, for any eigenstate $\bar H\ket{\phi_k} = E_k\ket{\phi_k}$, both $S$ and $V$ preserve the subspace spanned by the orthonormal states
	\begin{align}
	\ket{\varphi_k^0} &= \sum_j\beta_j\ket j\otimes\ket{\phi_k} \quad{\rm and}\\
	\ket{\varphi_k^1} &= \frac 1{\sqrt{1-E_k^2}}(V-E_k)\ket{\varphi_k^0},
	\end{align}
and it is easy to show that in the above basis
\begin{equation}
S = \left(
\begin{array}{cc} -1&0\\0&1 \end{array}
\right), \quad
V = \left(
\begin{array}{cc} E_k&\sqrt{1-E_k^2}\\\sqrt{1-E_k^2}&-E_k \end{array}
\right).
\end{equation}
We define the unitary walk operator $W = SVe^{i \pi}$ which has eigenvalues $e^{\pm i\theta_k}$ and eigenstates $\ket{\varphi_k^\pm} = (\ket{\varphi_k^0} \pm i \ket{\varphi_k^1})/\sqrt 2$, where $\cos\theta_k = E_k$. Thus, by preparing an initial state of the form $B\ket 0\otimes \ket{\tilde \phi_0}$ and performing phase estimation of $W$, we obtain either eigenvalue $\pm \theta_k$ with probability $|\bracket{\phi_k}{\tilde\phi_0}|^2$, and the post-measurement state is $\ket{\varphi_k^\pm}$. We can also use iterative phase estimation \cite{kitaev2002classical} which  simply uses an ancillary qubit $C$ prepare in the state $\ket +$, performs ${}^CW$, and measure the ancilla in the $\ket\pm$ basis, whose conditional probabilities are $p_{\pm|k} = \frac12(1\pm E_k)$. Repeating reveals $E_k$ and prepares one of the states $\ket{\varphi_k^\pm}$. 
	
	While the states $\ket{\varphi_k^\pm}$ are not eigenstates of $H$, we can nonetheless use them to compute expectation values. For instance, say we we want to estimate $\bra{\phi_k} \sigma\ket{\phi_k}$ for some multi-qubit Pauli operator $\sigma$. Observe that
	\begin{equation}
	\bra{\varphi_k^\pm} \sigma \ket{\varphi_k^\pm} 
	= \frac 12 \left(1+\frac{\Gamma_\sigma-E_k^2}{1-E_k^2}\right) \bra{\phi_k} \sigma \ket{\phi_k}
	\end{equation}
	where $\Gamma_\sigma = \sum_j |\beta_j|^2 (-1)^{\sigma\star P_j}$ and $\sigma\star P_j = 0$ when $\sigma$ and $P_j$ commute and $\sigma\star P_j=1$ when they anti-commute. Since $E_k$ is known from the state preparation and $\Gamma_\sigma$ can be computed efficiently, this gives us access to any static expectation values. 
	
	Alternatively, if we insist on preparing an eigenstate of $H$ rather than $\ket{\varphi_k^\pm}$, we have two solutions. First, ignoring the sign in the phase during phase estimation directly furnishes $\ket{\varphi_k^0}$ as the post-measurement state. Second, following \cite{MW05a}, we can simply apply $B^\dagger$ and measure if the control qubits are returned to the state $\ket 0$. It is straightforward to verify that this occurs with probability $\frac 12$ and that in case of a positive outcome the resulting state of the system qubits is $\ket{\phi_k}$. In the case of a negative outcome, the quantum computer is collapsed onto the state $B^{\dagger}\ket{\varphi_k^1}$. At this point we can apply $B$ and perform quantum phase estimation of $W$ which will randomly project onto $\ket{\varphi_k^\pm}$, and iterate. The probability of a positive outcome after $\ell$ iterations is $1-\frac 12^\ell$. 
	
	In either case, these procedures make the scheme fully compatible with the Zeno ground state preparation outlined above \cite{somma2008quantum}. Had we instead chosen to perform adiabatic evolution with the operator $W$ itself, viewed as a function of $g$, we would have had to worry about the spectral gap to the states orthogonal to the space spanned by the $\ket{\varphi_k^\pm}$. But by completing a deterministic projection as described in this paragraph, we are guaranteed to always remain in this invariant subspace.
	
	\medskip
	\noindent{\em Fewer Single-Qubit Rotations by Unary Encoding} ---
	The walk operator $W$ is composed of $V$ and $B$. The complexity of transformation $V$ stems from its multiple control, i.e. it applies a Pauli operator conditioned on one of $N$ possible value.  Each application of $V$ requires $\cO(N\log N)$ Toffoli gates. The complexity of $B$ stems from its dependency on $N$ real numbers $\beta_j$. So just by parameter counting, $B$ requires $\Omega(N)$ generic single-qubit gates. But this last argument breaks down when several of the $\beta_j$ are equal, as naturally occurs in condensed matter systems and lattice gauge theories. 
	
	Suppose the (rescaled) Hamiltonian takes the form
	\begin{equation}
	\bar H = |\beta_0|^2 I + \sum_{k=1}^K |\beta_k|^2 \sum_{j=1}^{N_k} P_j^k
	\label{eq:H2}
	\end{equation}
	such that $\sum_k N_k = N$ and $\sum_{k=0}^K |\beta_k|^2N_k = 1$, which is a special case of \eq{H} where the $|\beta_j|^2$ are restricted to $K$ distinct values. In lattice models with short-range interactions, $K$ is a small constant and all $N_k$ are roughly equal to the number of particles $n$ (deviations are caused by lattice boundaries). To reduce the number of generic single-qubit rotations used to implement $W$, we will increase the number of control qubits from $\log(N+1)$ to $N$, which we partitions into $K$ registers, with register $k$ of size $N_k$. The algorithm will proceed exactly as above, except that the control register will be encoded in a unary basis, i.e. $\ket{j} = \ket{00\ldots 0100\ldots 0}$ where the $1$ is at position $j$, and the state $\ket{j=0} = \ket{00\ldots 0}$. 
	
	To implement $B$ in this representation, we use a $K$-qubit rotation to prepare the state $\sum_k \sqrt{N_k}\beta_k \ket {m_k}$ where $m_1=1$ and $m_{k+1} = m_k+N_k$, and we use the unary representation. In other words, this state is a superposition of a $1$ in the first position of each register and requires $K$ generic single-qubit rotations. Then, using  $N$ $\sqrt{\rm SWAP}$ gates, we delocalize the $1$ in each register to obtain
	\begin{equation}
	\ket{\beta} = \beta_0\ket 0 + \sum_k \beta_k\sum_{j=0}^{N_k-1} \ket{m_k+j}.
	\end{equation}  
	 To achieve this, we  assume that $N_k$ is a power of $2$ and we simply apply the gate $\sqrt{\rm SWAP}$ in a binary-tree like structure, noting that $\sqrt{\rm SWAP}(\alpha \ket 0 +\beta \ket 1)\ket 0 = \alpha \ket{00} + \frac\beta{\sqrt 2}(\ket{10} + \ket{01})$. When $N_k$ is not a power of 2, we can simply pad by adding terms $P_j^k = I$ for $j>N_k$. The total number of $\sqrt{\rm SWAP}$ gates used to implement $B$ is thus upper bounded by $2N$. 
	
	The transformation $V$ is straightforward to implement in the unary representation. It consist in applying the product of control-$P_j^k$ where the control bit is the $j$th bit of the $k$'th control register. Because $P_j^k$ is a Pauli operator, control-$P_ j^k$ is a Clifford gate, so $V$ is entirely composed of Clifford gates. 
	
	Note that to perform quantum phase estimation, the circuit $W$ will generally need to be applied conditioned on some additional control register. To realize this, only the single-qubit rotations used to prepare the state $\sum_k \sqrt{N_k}\beta_k \ket{m_k}$ need to be controlled. The other gates do not need to be controlled because they act trivially in the absence of these initial single-qubit rotations.  To summarize, the original binary encoding uses $\log N$ control qubits, $\cO(N)$ generic single-qubit rotations, and $\cO(N\log N)$ third-level gates (Toffoli) while the unary encoding uses $N$ control qubits, $\cO(K)$ generic single-qubit rotations, and $\cO(N)$ third-level gates ($\sqrt{\rm SWAP}$). 
	
	\medskip
	
	\noindent{\em Lattices with Long-Range Couplings} --- The unary encoding  provides an advantage when there are only a few distinct coupling-strengths in the model, while the binary encoding is best when most couplings are different. Lattice models with long-range interaction provide an intermediate regime, where there are many distinct couplings, but each one is repeated an extensive number of times. For instance, a long-range 1D Ising interaction $J \sum_{i<j} \sigma_z^i\sigma_z^j/(i-j)^\alpha$ has $K  = n$ distinct coupling strengths $J_k = J/k^\alpha$, and each strength is repeated $n$ times, falling somewhere between the unary and binary encoding regimes.
		
	A good compromise is to prepare a $n$-qubit {\em coupling} state $\sum_k \sqrt n J_k \ket k$ to encode each coupling strength $J_k$ in unary representation, and use a $\log n$ qubit register in state $\frac 1{\sqrt n} \sum_i \ket i$ to encode the pair of sites $i$ and $i+k$ that are coupled in binary representation. The transformation $V$ can be realized by swapping qubits $i$ and $i+k$ into position 0 and 1, applying a pair of Pauli operators controlled by the state of the coupling qubit, and unswapping the two qubits to their original position. Swapping a qubit into a given position can be realized with $\cO(n)$ uses of the third-level gate control-SWAP, in a circuit of depth $n$ (or $\log n$ using $n/2$ additional ancillary qubits). Thus, this scheme requires $\cO(K)$ generic single-qubit rotations, $\cO(Kn)$ third-level gates and $\cO(K+\log n)$ ancillary qubits, which compares respectively to $\cO(Kn)$, $\cO(Kn\log( Kn))$, and $\cO(\log(Kn))$ for the binary encoding and $\cO(K)$, $\cO(Kn)$ and $\cO(Kn)$ for the unary encoding.
	
	\medskip
	
	\noindent{\em Discussion} --- We will now discuss the advantages and drawbacks of the methods we proposed above. The main drawback of using $W$ instead of the time-evolution operator is the rescaling $\cN \in \cO(N)$ of the spectral gap. To compensate for this rescaling, we will need to apply the operator $W$ a total of $ \cN t$ times to reach the same energy resolution that is achieved with $U(t)$. So the advantage of the proposed method depends on the number of Trotter steps that are needed to approximate $U(t)$ to some given accuracy. 
	
	Consider first a lattice model with short range interactions. The analysis of Appendix B in \cite{WBCH13a}  shows that the number of Trotter steps required to implement $U(t)$ to accuracy $\delta$ is $N_T \in \cO(t^{3/2}\sqrt{\frac n{\delta }})$. To resolve the ground state energy -- which is separated from the rest of the spectrum by an energy gap $\Delta$ -- we need $t\gg 1/\Delta$ and $\delta \ll \Delta$, which yields $N_T \in \cO(\frac{\sqrt n}{\Delta^{2}})$. Since each Trotter step requires $\cO(n)$ generic single-qubit rotations, the total cost of implementing an energy measurement capable of resolving the ground state is dominated by $\cO(\frac {n^{3/2}}{ \Delta^2})$ single-qubit rotations. On the other hand, the combination of schemes proposed here require $\cO(\frac \cN \Delta) = \cO(\frac n \Delta)$ single-qubit rotations and $\cO(n\frac{\cN} \Delta) = \cO(\frac{n^2}\Delta)$ third-level gates. Thus, our approach could offer important advantages at intermediate values of $n$, when the gate synthesis cost is dominant. We note that this analysis depends on the error criteria used. The Trotter error may scale differently  if one is interested not in the wave function but only in expectation values of local observables. 
	
	Second, consider a quantum chemistry problem. By the nature of Coulomb's force, once written in second quantized form $H = \sum_{pq} h_{pq} c_p^\dagger c_q + \sum_{pqrs} V_{pqrs} c_p^\dagger c_q^\dagger c_r c_s$, the Hamiltonian has $\cO(n^4)$ real parameters. Since every gate in a quantum circuit has a (small) constant number of free parameter, we clearly need  $O(n^4)$ generic single-qubit rotations to implement a single Trotter step $e^{-iH\delta}$.  The analysis of \cite{WBCH13a} shows that the number of Trotter steps required to implement $U(t)$ to accuracy $\delta$ is $N_T \in \cO(t^{3/2}\frac{n^5}{\sqrt \delta })$. The above reasoning thus leads to a total number of generic single-qubit rotations $\cO(\frac{n^{9}}{\Delta^2})$. The unary encoding is not of much help here, but using the transformation $W$ instead of $U(t)$ to perform an energy measurement requires a total of $\cO(n^4\frac{\cN}\Delta) = \cO(\frac{n^8}\Delta)$ single-qubit rotations. This is a substantial improvement over Trotter approaches.
	
	Note however that the previous paragraph was formulated in terms in upper bounds. It was found empirically that the number of Trotter steps can be much smaller in practice \cite{poulin2015trotter}. Indeed, for a range of small molecules, it was observed that the number of Trotter steps needs to scale as $N_T \in \cO(t^{3/2}\frac{n^\gamma}{\sqrt \delta })$ with $\gamma\sim 2.5$ instead of $\gamma=5$ revealed by the upper bound. This yields an overall scaling of $\cO(\frac{n^{\gamma+4}}{\Delta^{2}})$ to resolve the ground state by Trotter approximation. Using the same range of small molecules as used in \cite{poulin2015trotter}, we find that $\cN$ scales like $n^\eta$ with $\eta \sim 1.6$ as opposed to $\eta=4$ suggested by the upper bound. This yield an overall scaling of $\cO(\frac{n^{\eta+4}}\Delta)$ to resolve the ground state using our technique, leading roughly to the same savings by a factor $\frac n{{\Delta}}$ as derived from the rigorous upper bounds above. It will be interesting to compare both approaches for large molecules -- which will be possible once we have quantum computers with a large number of logical qubits.
	
Taylor expansions schemes  \cite{berry2015simulating} rely on breaking up the time-evolution operator $U(t) = U(t/r)^r$ into $r$ segments, and approximating each segment by a Taylor series 
\begin{equation}
U(t/r) \approx \sum_{m=0}^M \frac 1{m!}(-iHt/r)^m.
\label{eq:Taylor}
\end{equation}
To resolve the ground state, $r$ needs to be greater than $\frac {\|H\|} \Delta$, and $M\sim \log( \frac {\|H\|}{\Delta^2})$. The scheme then proceeds by implementing a unitary transformation $W$ similar to the one constructed here, but with the sum of \eq{Taylor} replacing the Hamiltonian. While the Hamiltonian has $N$ terms, this sum has $\sim NM$ terms, so requires $M$ times more gates. So in summary, our scheme has all the advantages of the Taylor series, except that it only needs a first order $M=1$ expansion and uses a single segment, thus requires $Mr \sim \frac {\|H\|} \Delta \log( \frac {\|H\|}{\Delta^2})$ times fewer gates.

We see that the main cost of using $W$ stems from the rescaling factor $\cN$, which is equal to the sum of the absolute value of the terms of the Hamiltonian. To improve the scaling, we need a different function $f(H)$ with a greater slope near $E_0$. For instance, the Taylor series and quantum signal processing approaches~\cite{low2017usa} allow us to synthesize functions where $f(H)$ are degree-$M$ polynomials. However, the reduction in $\cN$ comes with an multiplicative cost $\mathcal{O}(M)$.  We leave this problem open for future research.

In conclusion, we have presented two techniques to simplify energy measurements on a quantum computer. The general principle behind the first technique is to simulate not imitate: implement physically inaccessible transformations to improve simulation algorithms. The second technique uses special features of lattice-based Hamiltonians, namely the small number of independent parameters due to translational invariance. This illustrates the importance for quantum information scientists to work alongside domain experts in quantum chemistry, condensed matter, high energy, etc. to find other features of the models that can be exploited to further improve quantum simulations. Finally, we note that the ideas presented here extend beyond simulations and can be directly applied to spectral measurement of other operators such as used in quantum algorithms for linear systems of equations \cite{HHL09a} for instance.
	
		\medskip
	\noindent{\em Acknowledgements} We thank Thomas H\"aner for interesting discussions and Sergey Bravyi for feedback, and acknowledge support by the Swiss National Science Foundation, the NCCR QSIT, Canada's NSERC, Caltech's IQIM, and the Simons Foundation.
	

\begin{thebibliography}{41}%
\makeatletter
\providecommand \@ifxundefined [1]{%
 \@ifx{#1\undefined}
}%
\providecommand \@ifnum [1]{%
 \ifnum #1\expandafter \@firstoftwo
 \else \expandafter \@secondoftwo
 \fi
}%
\providecommand \@ifx [1]{%
 \ifx #1\expandafter \@firstoftwo
 \else \expandafter \@secondoftwo
 \fi
}%
\providecommand \natexlab [1]{#1}%
\providecommand \enquote  [1]{``#1''}%
\providecommand \bibnamefont  [1]{#1}%
\providecommand \bibfnamefont [1]{#1}%
\providecommand \citenamefont [1]{#1}%
\providecommand \href@noop [0]{\@secondoftwo}%
\providecommand \href [0]{\begingroup \@sanitize@url \@href}%
\providecommand \@href[1]{\@@startlink{#1}\@@href}%
\providecommand \@@href[1]{\endgroup#1\@@endlink}%
\providecommand \@sanitize@url [0]{\catcode `\\12\catcode `\$12\catcode
  `\&12\catcode `\#12\catcode `\^12\catcode `\_12\catcode `\%12\relax}%
\providecommand \@@startlink[1]{}%
\providecommand \@@endlink[0]{}%
\providecommand \url  [0]{\begingroup\@sanitize@url \@url }%
\providecommand \@url [1]{\endgroup\@href {#1}{\urlprefix }}%
\providecommand \urlprefix  [0]{URL }%
\providecommand \Eprint [0]{\href }%
\@ifxundefined \urlstyle {%
  \providecommand \doi  [0]{\begingroup \@sanitize@url \@doi}%
  \providecommand \@doi [1]{\endgroup \@@startlink {\doibase
  #1}doi:\discretionary {}{}{}#1\@@endlink }%
}{%
  \providecommand \doi  [0]{doi:\discretionary{}{}{}\begingroup
  \urlstyle{rm}\Url }%
}%
\providecommand \doibase [0]{http://dx.doi.org/}%
\providecommand \Doi [0]{\begingroup \@sanitize@url \@Doi }%
\providecommand \@Doi  [1]{\endgroup\@@startlink{\doibase#1}\@@Doi}%
\providecommand \@@Doi [1]{#1\@@endlink}%
\providecommand \selectlanguage [0]{\@gobble}%
\providecommand \bibinfo  [0]{\@secondoftwo}%
\providecommand \bibfield  [0]{\@secondoftwo}%
\providecommand \translation [1]{[#1]}%
\providecommand \BibitemOpen [0]{}%
\providecommand \bibitemStop [0]{}%
\providecommand \bibitemNoStop [0]{.\EOS\space}%
\providecommand \EOS [0]{\spacefactor3000\relax}%
\providecommand \BibitemShut  [1]{\csname bibitem#1\endcsname}%
\bibitem [{\citenamefont {Feynman}(1982)}]{feynman1982simulating}%
  \BibitemOpen
  \bibfield  {author} {\bibinfo {author} {\bibfnamefont {Richard~P.}\
  \bibnamefont {Feynman}},\ }\bibfield  {title} {\enquote {\bibinfo {title}
  {Simulating physics with computers},}\ }\Doi {10.1007/BF02650179} {\bibfield
  {journal} {\bibinfo  {journal} {International Journal of Theoretical
  Physics},\ }\textbf {\bibinfo {volume} {21}},\ \bibinfo {pages} {467--488}
  (\bibinfo {year} {1982})}\BibitemShut {NoStop}%
\bibitem [{\citenamefont {Lloyd}(1996)}]{lloyd1996universal}%
  \BibitemOpen
  \bibfield  {author} {\bibinfo {author} {\bibfnamefont {Seth}\ \bibnamefont
  {Lloyd}},\ }\bibfield  {title} {\enquote {\bibinfo {title} {Universal quantum
  simulators},}\ }\Doi {10.1126/science.273.5278.1073} {\bibfield  {journal}
  {\bibinfo  {journal} {Science},\ }\textbf {\bibinfo {volume} {273}},\
  \bibinfo {pages} {1073--1077} (\bibinfo {year} {1996})}\BibitemShut {NoStop}%
\bibitem [{\citenamefont {Abrams}\ and\ \citenamefont
  {Lloyd}(1997)}]{abrams1997simulation}%
  \BibitemOpen
  \bibfield  {author} {\bibinfo {author} {\bibfnamefont {Daniel~S.}\
  \bibnamefont {Abrams}}\ and\ \bibinfo {author} {\bibfnamefont {Seth}\
  \bibnamefont {Lloyd}},\ }\bibfield  {title} {\enquote {\bibinfo {title}
  {Simulation of many-body {F}ermi systems on a universal quantum computer},}\
  }\Doi {10.1103/PhysRevLett.79.2586} {\bibfield  {journal} {\bibinfo
  {journal} {Physical Review Letters},\ }\textbf {\bibinfo {volume} {79}},\
  \bibinfo {pages} {2586} (\bibinfo {year} {1997})}\BibitemShut {NoStop}%
\bibitem [{\citenamefont {Somma}\ \emph {et~al.}(2002)\citenamefont {Somma},
  \citenamefont {Ortiz}, \citenamefont {Gubernatis}, \citenamefont {Knill},\
  and\ \citenamefont {Laflamme}}]{gubernatis}%
  \BibitemOpen
  \bibfield  {author} {\bibinfo {author} {\bibfnamefont {R.}~\bibnamefont
  {Somma}}, \bibinfo {author} {\bibfnamefont {G.}~\bibnamefont {Ortiz}},
  \bibinfo {author} {\bibfnamefont {J.~E.}\ \bibnamefont {Gubernatis}},
  \bibinfo {author} {\bibfnamefont {E.}~\bibnamefont {Knill}}, \ and\ \bibinfo
  {author} {\bibfnamefont {R.}~\bibnamefont {Laflamme}},\ }\bibfield  {title}
  {\enquote {\bibinfo {title} {Simulating physical phenomena by quantum
  networks},}\ }\Doi {10.1103/PhysRevA.65.042323} {\bibfield  {journal}
  {\bibinfo  {journal} {Physical Review A},\ }\textbf {\bibinfo {volume}
  {65}},\ \bibinfo {pages} {042323} (\bibinfo {year} {2002})}\BibitemShut
  {NoStop}%
\bibitem [{\citenamefont {Bauer}\ \emph {et~al.}(2016)\citenamefont {Bauer},
  \citenamefont {Wecker}, \citenamefont {Millis}, \citenamefont {Hastings},\
  and\ \citenamefont {Troyer}}]{bauer2016hybrid}%
  \BibitemOpen
  \bibfield  {author} {\bibinfo {author} {\bibfnamefont {Bela}\ \bibnamefont
  {Bauer}}, \bibinfo {author} {\bibfnamefont {Dave}\ \bibnamefont {Wecker}},
  \bibinfo {author} {\bibfnamefont {Andrew~J.}\ \bibnamefont {Millis}},
  \bibinfo {author} {\bibfnamefont {Matthew~B.}\ \bibnamefont {Hastings}}, \
  and\ \bibinfo {author} {\bibfnamefont {Matthias}\ \bibnamefont {Troyer}},\
  }\bibfield  {title} {\enquote {\bibinfo {title} {Hybrid quantum-classical
  approach to correlated materials},}\ }\Doi {10.1103/PhysRevX.6.031045}
  {\bibfield  {journal} {\bibinfo  {journal} {Physical Review X},\ }\textbf
  {\bibinfo {volume} {6}},\ \bibinfo {pages} {031045} (\bibinfo {year}
  {2016})}\BibitemShut {NoStop}%
\bibitem [{\citenamefont {Wecker}\ \emph {et~al.}(2015)\citenamefont {Wecker},
  \citenamefont {Hastings}, \citenamefont {Wiebe}, \citenamefont {Clark},
  \citenamefont {Nayak},\ and\ \citenamefont {Troyer}}]{wecker2015solving}%
  \BibitemOpen
  \bibfield  {author} {\bibinfo {author} {\bibfnamefont {Dave}\ \bibnamefont
  {Wecker}}, \bibinfo {author} {\bibfnamefont {Matthew~B.}\ \bibnamefont
  {Hastings}}, \bibinfo {author} {\bibfnamefont {Nathan}\ \bibnamefont
  {Wiebe}}, \bibinfo {author} {\bibfnamefont {Bryan~K.}\ \bibnamefont {Clark}},
  \bibinfo {author} {\bibfnamefont {Chetan}\ \bibnamefont {Nayak}}, \ and\
  \bibinfo {author} {\bibfnamefont {Matthias}\ \bibnamefont {Troyer}},\
  }\bibfield  {title} {\enquote {\bibinfo {title} {Solving strongly correlated
  electron models on a quantum computer},}\ }\Doi {10.1103/PhysRevA.92.062318}
  {\bibfield  {journal} {\bibinfo  {journal} {Physical Review A},\ }\textbf
  {\bibinfo {volume} {92}},\ \bibinfo {pages} {062318} (\bibinfo {year}
  {2015})}\BibitemShut {NoStop}%
\bibitem [{\citenamefont {Aspuru-Guzik}\ \emph {et~al.}(2005)\citenamefont
  {Aspuru-Guzik}, \citenamefont {Dutoi}, \citenamefont {Love},\ and\
  \citenamefont {Head-Gordon}}]{aspuru2005simulated}%
  \BibitemOpen
  \bibfield  {author} {\bibinfo {author} {\bibfnamefont {Al{\'a}n}\
  \bibnamefont {Aspuru-Guzik}}, \bibinfo {author} {\bibfnamefont {Anthony~D.}\
  \bibnamefont {Dutoi}}, \bibinfo {author} {\bibfnamefont {Peter~J.}\
  \bibnamefont {Love}}, \ and\ \bibinfo {author} {\bibfnamefont {Martin}\
  \bibnamefont {Head-Gordon}},\ }\bibfield  {title} {\enquote {\bibinfo {title}
  {Simulated quantum computation of molecular energies},}\ }\Doi
  {10.1126/science.1113479} {\bibfield  {journal} {\bibinfo  {journal}
  {Science},\ }\textbf {\bibinfo {volume} {309}},\ \bibinfo {pages}
  {1704--1707} (\bibinfo {year} {2005})}\BibitemShut {NoStop}%
\bibitem [{\citenamefont {Jordan}\ \emph {et~al.}(2012)\citenamefont {Jordan},
  \citenamefont {Lee},\ and\ \citenamefont {Preskill}}]{jordan2012quantum}%
  \BibitemOpen
  \bibfield  {author} {\bibinfo {author} {\bibfnamefont {Stephen~P.}\
  \bibnamefont {Jordan}}, \bibinfo {author} {\bibfnamefont {Keith S.~M.}\
  \bibnamefont {Lee}}, \ and\ \bibinfo {author} {\bibfnamefont {John}\
  \bibnamefont {Preskill}},\ }\bibfield  {title} {\enquote {\bibinfo {title}
  {Quantum algorithms for quantum field theories},}\ }\Doi
  {10.1126/science.1217069} {\bibfield  {journal} {\bibinfo  {journal}
  {Science},\ }\textbf {\bibinfo {volume} {336}},\ \bibinfo {pages}
  {1130--1133} (\bibinfo {year} {2012})}\BibitemShut {NoStop}%
\bibitem [{\citenamefont {Berry}\ \emph {et~al.}(2015)\citenamefont {Berry},
  \citenamefont {Childs}, \citenamefont {Cleve}, \citenamefont {Kothari},\ and\
  \citenamefont {Somma}}]{berry2015simulating}%
  \BibitemOpen
  \bibfield  {author} {\bibinfo {author} {\bibfnamefont {Dominic~W.}\
  \bibnamefont {Berry}}, \bibinfo {author} {\bibfnamefont {Andrew~M.}\
  \bibnamefont {Childs}}, \bibinfo {author} {\bibfnamefont {Richard}\
  \bibnamefont {Cleve}}, \bibinfo {author} {\bibfnamefont {Robin}\ \bibnamefont
  {Kothari}}, \ and\ \bibinfo {author} {\bibfnamefont {Rolando~D.}\
  \bibnamefont {Somma}},\ }\bibfield  {title} {\enquote {\bibinfo {title}
  {Simulating hamiltonian dynamics with a truncated taylor series},}\ }\Doi
  {10.1103/PhysRevLett.114.090502} {\bibfield  {journal} {\bibinfo  {journal}
  {Physical Review Letters},\ }\textbf {\bibinfo {volume} {114}},\ \bibinfo
  {pages} {090502} (\bibinfo {year} {2015})}\BibitemShut {NoStop}%
\bibitem [{\citenamefont {Low}\ and\ \citenamefont
  {Chuang}(2017){\natexlab{a}}}]{low2017optimal}%
  \BibitemOpen
  \bibfield  {author} {\bibinfo {author} {\bibfnamefont {Guang~Hao}\
  \bibnamefont {Low}}\ and\ \bibinfo {author} {\bibfnamefont {Isaac~L.}\
  \bibnamefont {Chuang}},\ }\bibfield  {title} {\enquote {\bibinfo {title}
  {Optimal hamiltonian simulation by quantum signal processing},}\ }\Doi
  {10.1103/PhysRevLett.118.010501} {\bibfield  {journal} {\bibinfo  {journal}
  {Physical Review Letters},\ }\textbf {\bibinfo {volume} {118}},\ \bibinfo
  {pages} {010501} (\bibinfo {year} {2017}{\natexlab{a}})}\BibitemShut
  {NoStop}%
\bibitem [{\citenamefont {Kitaev}(1995)}]{kitaev1995quantum}%
  \BibitemOpen
  \bibfield  {author} {\bibinfo {author} {\bibfnamefont {A.~Yu.}\ \bibnamefont
  {Kitaev}},\ }\bibfield  {title} {\enquote {\bibinfo {title} {Quantum
  measurements and the abelian stabilizer problem},}\ }\href
  {https://arxiv.org/abs/quant-ph/9511026} {\bibfield  {journal} {\bibinfo
  {journal} {arXiv preprint quant-ph/9511026}} (\bibinfo {year}
  {1995})}\BibitemShut {NoStop}%
\bibitem [{\citenamefont {Cleve}\ \emph {et~al.}(1998)\citenamefont {Cleve},
  \citenamefont {Ekert}, \citenamefont {Macchiavello},\ and\ \citenamefont
  {Mosca}}]{cleve1998quantum}%
  \BibitemOpen
  \bibfield  {author} {\bibinfo {author} {\bibfnamefont {Richard}\ \bibnamefont
  {Cleve}}, \bibinfo {author} {\bibfnamefont {Artur}\ \bibnamefont {Ekert}},
  \bibinfo {author} {\bibfnamefont {Chiara}\ \bibnamefont {Macchiavello}}, \
  and\ \bibinfo {author} {\bibfnamefont {Michele}\ \bibnamefont {Mosca}},\
  }\bibfield  {title} {\enquote {\bibinfo {title} {Quantum algorithms
  revisited},}\ }in\ \Doi {10.1098/rspa.1998.0164} {\emph {\bibinfo {booktitle}
  {Proceedings of the Royal Society of London A: Mathematical, Physical and
  Engineering Sciences}}},\ Vol.\ \bibinfo {volume} {454}\ (\bibinfo
  {organization} {The Royal Society},\ \bibinfo {year} {1998})\ pp.\ \bibinfo
  {pages} {339--354}\BibitemShut {NoStop}%
\bibitem [{\citenamefont {Abrams}\ and\ \citenamefont
  {Lloyd}(1999)}]{abrams1999quantum}%
  \BibitemOpen
  \bibfield  {author} {\bibinfo {author} {\bibfnamefont {Daniel~S.}\
  \bibnamefont {Abrams}}\ and\ \bibinfo {author} {\bibfnamefont {Seth}\
  \bibnamefont {Lloyd}},\ }\bibfield  {title} {\enquote {\bibinfo {title}
  {Quantum algorithm providing exponential speed increase for finding
  eigenvalues and eigenvectors},}\ }\Doi {10.1103/PhysRevLett.83.5162}
  {\bibfield  {journal} {\bibinfo  {journal} {Physical Review Letters},\
  }\textbf {\bibinfo {volume} {83}},\ \bibinfo {pages} {5162} (\bibinfo {year}
  {1999})}\BibitemShut {NoStop}%
\bibitem [{\citenamefont {Somma}\ \emph {et~al.}(2008)\citenamefont {Somma},
  \citenamefont {Boixo}, \citenamefont {Barnum},\ and\ \citenamefont
  {Knill}}]{somma2008quantum}%
  \BibitemOpen
  \bibfield  {author} {\bibinfo {author} {\bibfnamefont {R.~D.}\ \bibnamefont
  {Somma}}, \bibinfo {author} {\bibfnamefont {S.}~\bibnamefont {Boixo}},
  \bibinfo {author} {\bibfnamefont {Howard}\ \bibnamefont {Barnum}}, \ and\
  \bibinfo {author} {\bibfnamefont {E.}~\bibnamefont {Knill}},\ }\bibfield
  {title} {\enquote {\bibinfo {title} {Quantum simulations of classical
  annealing processes},}\ }\Doi {10.1103/PhysRevLett.101.130504} {\bibfield
  {journal} {\bibinfo  {journal} {Physical Review Letters},\ }\textbf {\bibinfo
  {volume} {101}},\ \bibinfo {pages} {130504} (\bibinfo {year}
  {2008})}\BibitemShut {NoStop}%
\bibitem [{\citenamefont {Farhi}\ \emph {et~al.}(2000)\citenamefont {Farhi},
  \citenamefont {Goldstone}, \citenamefont {Gutmann},\ and\ \citenamefont
  {Sipser}}]{FGGS00a}%
  \BibitemOpen
  \bibfield  {author} {\bibinfo {author} {\bibfnamefont {Edward}\ \bibnamefont
  {Farhi}}, \bibinfo {author} {\bibfnamefont {Jeffrey}\ \bibnamefont
  {Goldstone}}, \bibinfo {author} {\bibfnamefont {Sam}\ \bibnamefont
  {Gutmann}}, \ and\ \bibinfo {author} {\bibfnamefont {Michael}\ \bibnamefont
  {Sipser}},\ }\href@noop {} {\enquote {\bibinfo {title} {Quantum computation
  by adiabatic evolution},}\ } (\bibinfo {year} {2000}),\ \Eprint
  {http://arxiv.org/abs/quant-ph/0001106} {quant-ph/0001106} \BibitemShut
  {NoStop}%
\bibitem [{\citenamefont {Aharonov}\ and\ \citenamefont
  {Ta-Shma}(2003)}]{AT03b}%
  \BibitemOpen
  \bibfield  {author} {\bibinfo {author} {\bibfnamefont {D.}~\bibnamefont
  {Aharonov}}\ and\ \bibinfo {author} {\bibfnamefont {A.}~\bibnamefont
  {Ta-Shma}},\ }\bibfield  {title} {\enquote {\bibinfo {title} {Adiabatic
  quantum state generation and statistical zero knowledge},}\ }\href@noop {}
  {\bibfield  {journal} {\bibinfo  {journal} {Proc. 35th Annual ACM Symp. on
  Theo. Comp.},\ \bibinfo {pages} {20}} (\bibinfo {year} {2003})}\BibitemShut
  {NoStop}%
\bibitem [{\citenamefont {Temme}\ \emph {et~al.}(2011)\citenamefont {Temme},
  \citenamefont {Osborne}, \citenamefont {Vollbrecht}, \citenamefont {Poulin},\
  and\ \citenamefont {Verstraete}}]{temme2011quantum}%
  \BibitemOpen
  \bibfield  {author} {\bibinfo {author} {\bibfnamefont {Kristan}\ \bibnamefont
  {Temme}}, \bibinfo {author} {\bibfnamefont {Tobias~J.}\ \bibnamefont
  {Osborne}}, \bibinfo {author} {\bibfnamefont {Karl~G.}\ \bibnamefont
  {Vollbrecht}}, \bibinfo {author} {\bibfnamefont {David}\ \bibnamefont
  {Poulin}}, \ and\ \bibinfo {author} {\bibfnamefont {Frank}\ \bibnamefont
  {Verstraete}},\ }\bibfield  {title} {\enquote {\bibinfo {title} {Quantum
  metropolis sampling},}\ }\Doi {10.1038/nature09770} {\bibfield  {journal}
  {\bibinfo  {journal} {Nature},\ }\textbf {\bibinfo {volume} {471}},\ \bibinfo
  {pages} {87--90} (\bibinfo {year} {2011})}\BibitemShut {NoStop}%
\bibitem [{\citenamefont {Yung}\ and\ \citenamefont
  {Aspuru-Guzik}(2012)}]{yung2012quantum}%
  \BibitemOpen
  \bibfield  {author} {\bibinfo {author} {\bibfnamefont {Man-Hong}\
  \bibnamefont {Yung}}\ and\ \bibinfo {author} {\bibfnamefont {Al{\'a}n}\
  \bibnamefont {Aspuru-Guzik}},\ }\bibfield  {title} {\enquote {\bibinfo
  {title} {A quantum--quantum metropolis algorithm},}\ }\Doi
  {10.1073/pnas.1111758109} {\bibfield  {journal} {\bibinfo  {journal}
  {Proceedings of the National Academy of Sciences},\ }\textbf {\bibinfo
  {volume} {109}},\ \bibinfo {pages} {754--759} (\bibinfo {year}
  {2012})}\BibitemShut {NoStop}%
\bibitem [{\citenamefont {Poulin}\ and\ \citenamefont
  {Wocjan}(2009)}]{poulin2009preparing}%
  \BibitemOpen
  \bibfield  {author} {\bibinfo {author} {\bibfnamefont {David}\ \bibnamefont
  {Poulin}}\ and\ \bibinfo {author} {\bibfnamefont {Pawel}\ \bibnamefont
  {Wocjan}},\ }\bibfield  {title} {\enquote {\bibinfo {title} {Preparing ground
  states of quantum many-body systems on a quantum computer},}\ }\Doi
  {10.1103/PhysRevLett.102.130503} {\bibfield  {journal} {\bibinfo  {journal}
  {Physical Review Letters},\ }\textbf {\bibinfo {volume} {102}},\ \bibinfo
  {pages} {130503} (\bibinfo {year} {2009})}\BibitemShut {NoStop}%
\bibitem [{\citenamefont {Szegedy}(2004)}]{szegedy2004quantum}%
  \BibitemOpen
  \bibfield  {author} {\bibinfo {author} {\bibfnamefont {Mario}\ \bibnamefont
  {Szegedy}},\ }\bibfield  {title} {\enquote {\bibinfo {title} {Quantum
  speed-up of {M}arkov chain based algorithms},}\ }in\ \Doi
  {10.1109/FOCS.2004.53} {\emph {\bibinfo {booktitle} {Foundations of Computer
  Science, 2004. Proceedings. 45th Annual IEEE Symposium on}}}\ (\bibinfo
  {organization} {IEEE},\ \bibinfo {year} {2004})\ pp.\ \bibinfo {pages}
  {32--41}\BibitemShut {NoStop}%
\bibitem [{\citenamefont {Low}\ and\ \citenamefont
  {Chuang}(2016)}]{low2016qubitization}%
  \BibitemOpen
  \bibfield  {author} {\bibinfo {author} {\bibfnamefont {Guang~Hao}\
  \bibnamefont {Low}}\ and\ \bibinfo {author} {\bibfnamefont {Isaac~L.}\
  \bibnamefont {Chuang}},\ }\bibfield  {title} {\enquote {\bibinfo {title}
  {Hamiltonian simulation by {Q}ubitization},}\ }\href
  {https://arxiv.org/abs/1610.06546} {\bibfield  {journal} {\bibinfo  {journal}
  {arXiv preprint arXiv:1610.06546}} (\bibinfo {year} {2016})}\BibitemShut
  {NoStop}%
\bibitem [{\citenamefont {Aharonov}\ and\ \citenamefont
  {Ben-Or}(1997)}]{aharonov1997fault}%
  \BibitemOpen
  \bibfield  {author} {\bibinfo {author} {\bibfnamefont {Dorit}\ \bibnamefont
  {Aharonov}}\ and\ \bibinfo {author} {\bibfnamefont {Michael}\ \bibnamefont
  {Ben-Or}},\ }\bibfield  {title} {\enquote {\bibinfo {title} {Fault-tolerant
  quantum computation with constant error},}\ }in\ \Doi {10.1145/258533.258579}
  {\emph {\bibinfo {booktitle} {Proceedings of the twenty-ninth annual ACM
  symposium on Theory of computing}}}\ (\bibinfo {organization} {ACM},\
  \bibinfo {year} {1997})\ pp.\ \bibinfo {pages} {176--188}\BibitemShut
  {NoStop}%
\bibitem [{\citenamefont {Shor}(1996)}]{shor1996fault}%
  \BibitemOpen
  \bibfield  {author} {\bibinfo {author} {\bibfnamefont {Peter~W.}\
  \bibnamefont {Shor}},\ }\bibfield  {title} {\enquote {\bibinfo {title}
  {Fault-tolerant quantum computation},}\ }in\ \Doi {10.1109/SFCS.1996.548464}
  {\emph {\bibinfo {booktitle} {Foundations of Computer Science, 1996.
  Proceedings., 37th Annual Symposium on}}}\ (\bibinfo {organization} {IEEE},\
  \bibinfo {year} {1996})\ pp.\ \bibinfo {pages} {56--65}\BibitemShut {NoStop}%
\bibitem [{\citenamefont {Steane}(1999)}]{steane1999efficient}%
  \BibitemOpen
  \bibfield  {author} {\bibinfo {author} {\bibfnamefont {Andrew~M.}\
  \bibnamefont {Steane}},\ }\bibfield  {title} {\enquote {\bibinfo {title}
  {Efficient fault-tolerant quantum computing},}\ }\Doi {10.1038/20127}
  {\bibfield  {journal} {\bibinfo  {journal} {Nature},\ }\textbf {\bibinfo
  {volume} {399}},\ \bibinfo {pages} {124--126} (\bibinfo {year}
  {1999})}\BibitemShut {NoStop}%
\bibitem [{\citenamefont {Raussendorf}\ and\ \citenamefont
  {Harrington}(2007)}]{raussendorf2007fault}%
  \BibitemOpen
  \bibfield  {author} {\bibinfo {author} {\bibfnamefont {Robert}\ \bibnamefont
  {Raussendorf}}\ and\ \bibinfo {author} {\bibfnamefont {Jim}\ \bibnamefont
  {Harrington}},\ }\bibfield  {title} {\enquote {\bibinfo {title}
  {Fault-tolerant quantum computation with high threshold in two dimensions},}\
  }\Doi {10.1103/PhysRevLett.98.190504} {\bibfield  {journal} {\bibinfo
  {journal} {Physical Review Letters},\ }\textbf {\bibinfo {volume} {98}},\
  \bibinfo {pages} {190504} (\bibinfo {year} {2007})}\BibitemShut {NoStop}%
\bibitem [{\citenamefont {Knill}(2004)}]{knill2004fault}%
  \BibitemOpen
  \bibfield  {author} {\bibinfo {author} {\bibfnamefont {Emanuel}\ \bibnamefont
  {Knill}},\ }\bibfield  {title} {\enquote {\bibinfo {title} {Fault-tolerant
  postselected quantum computation: Schemes},}\ }\href
  {https://arxiv.org/abs/quant-ph/0402171} {\bibfield  {journal} {\bibinfo
  {journal} {arXiv preprint quant-ph/0402171}} (\bibinfo {year}
  {2004})}\BibitemShut {NoStop}%
\bibitem [{\citenamefont {Bravyi}\ and\ \citenamefont
  {Kitaev}(2005)}]{bravyi2005universal}%
  \BibitemOpen
  \bibfield  {author} {\bibinfo {author} {\bibfnamefont {Sergey}\ \bibnamefont
  {Bravyi}}\ and\ \bibinfo {author} {\bibfnamefont {Alexei}\ \bibnamefont
  {Kitaev}},\ }\bibfield  {title} {\enquote {\bibinfo {title} {Universal
  quantum computation with ideal clifford gates and noisy ancillas},}\ }\Doi
  {10.1103/PhysRevA.71.022316} {\bibfield  {journal} {\bibinfo  {journal}
  {Physical Review A},\ }\textbf {\bibinfo {volume} {71}},\ \bibinfo {pages}
  {022316} (\bibinfo {year} {2005})}\BibitemShut {NoStop}%
\bibitem [{\citenamefont {Jones}(2013)}]{jones2013multilevel}%
  \BibitemOpen
  \bibfield  {author} {\bibinfo {author} {\bibfnamefont {Cody}\ \bibnamefont
  {Jones}},\ }\bibfield  {title} {\enquote {\bibinfo {title} {Multilevel
  distillation of magic states for quantum computing},}\ }\Doi
  {10.1103/PhysRevA.87.042305} {\bibfield  {journal} {\bibinfo  {journal}
  {Physical Review A},\ }\textbf {\bibinfo {volume} {87}},\ \bibinfo {pages}
  {042305} (\bibinfo {year} {2013})}\BibitemShut {NoStop}%
\bibitem [{\citenamefont {Haah}\ \emph {et~al.}(2017)\citenamefont {Haah},
  \citenamefont {Hastings}, \citenamefont {Poulin},\ and\ \citenamefont
  {Wecker}}]{haah2017magic}%
  \BibitemOpen
  \bibfield  {author} {\bibinfo {author} {\bibfnamefont {Jeongwan}\
  \bibnamefont {Haah}}, \bibinfo {author} {\bibfnamefont {Matthew~B.}\
  \bibnamefont {Hastings}}, \bibinfo {author} {\bibfnamefont {D.}~\bibnamefont
  {Poulin}}, \ and\ \bibinfo {author} {\bibfnamefont {D.}~\bibnamefont
  {Wecker}},\ }\bibfield  {title} {\enquote {\bibinfo {title} {Magic state
  distillation with low space overhead and optimal asymptotic input count},}\
  }\Doi {10.22331/q-2017-10-03-31} {\bibfield  {journal} {\bibinfo  {journal}
  {Quantum},\ }\textbf {\bibinfo {volume} {1}},\ \bibinfo {pages} {31}
  (\bibinfo {year} {2017})}\BibitemShut {NoStop}%
\bibitem [{\citenamefont {Hastings}\ and\ \citenamefont
  {Haah}(2017)}]{hastings2017distillation}%
  \BibitemOpen
  \bibfield  {author} {\bibinfo {author} {\bibfnamefont {Matthew~B.}\
  \bibnamefont {Hastings}}\ and\ \bibinfo {author} {\bibfnamefont {Jeongwan}\
  \bibnamefont {Haah}},\ }\bibfield  {title} {\enquote {\bibinfo {title}
  {Distillation with sublogarithmic overhead},}\ }\href
  {https://arxiv.org/abs/1709.03543} {\bibfield  {journal} {\bibinfo  {journal}
  {arXiv preprint arXiv:1709.03543}} (\bibinfo {year} {2017})}\BibitemShut
  {NoStop}%
\bibitem [{\citenamefont {Gottesman}\ and\ \citenamefont
  {Chuang}(1999)}]{gottesman1999demonstrating}%
  \BibitemOpen
  \bibfield  {author} {\bibinfo {author} {\bibfnamefont {Daniel}\ \bibnamefont
  {Gottesman}}\ and\ \bibinfo {author} {\bibfnamefont {Isaac~L.}\ \bibnamefont
  {Chuang}},\ }\bibfield  {title} {\enquote {\bibinfo {title} {Demonstrating
  the viability of universal quantum computation using teleportation and
  single-qubit operations},}\ }\Doi {10.1038/46503} {\bibfield  {journal}
  {\bibinfo  {journal} {Nature},\ }\textbf {\bibinfo {volume} {402}},\ \bibinfo
  {pages} {390--393} (\bibinfo {year} {1999})}\BibitemShut {NoStop}%
\bibitem [{\citenamefont {Dawson}\ and\ \citenamefont
  {Nielsen}(2006)}]{dawson2005solovay}%
  \BibitemOpen
  \bibfield  {author} {\bibinfo {author} {\bibfnamefont {Christopher~M.}\
  \bibnamefont {Dawson}}\ and\ \bibinfo {author} {\bibfnamefont {Michael~A.}\
  \bibnamefont {Nielsen}},\ }\bibfield  {title} {\enquote {\bibinfo {title}
  {The {S}olovay-{K}itaev algorithm},}\ }\href
  {https://dl.acm.org/citation.cfm?id=2011685} {\bibfield  {journal} {\bibinfo
  {journal} {Quantum Information {\&} Computation},\ }\textbf {\bibinfo
  {volume} {6}},\ \bibinfo {pages} {81--95} (\bibinfo {year}
  {2006})}\BibitemShut {NoStop}%
\bibitem [{\citenamefont {Kliuchnikov}\ \emph {et~al.}(2013)\citenamefont
  {Kliuchnikov}, \citenamefont {Maslov},\ and\ \citenamefont {Mosca}}]{KMM12a}%
  \BibitemOpen
  \bibfield  {author} {\bibinfo {author} {\bibfnamefont {V.}~\bibnamefont
  {Kliuchnikov}}, \bibinfo {author} {\bibfnamefont {D.}~\bibnamefont {Maslov}},
  \ and\ \bibinfo {author} {\bibfnamefont {M.}~\bibnamefont {Mosca}},\
  }\bibfield  {title} {\enquote {\bibinfo {title} {Asymptotically optimal
  approximation of single qubit unitaries by {C}lifford and {T} circuits using
  a constant number of ancillary qubits},}\ }\Doi
  {10.1103/PhysRevLett.110.190502} {\bibfield  {journal} {\bibinfo  {journal}
  {Physical Review Letters},\ }\textbf {\bibinfo {volume} {110}},\ \bibinfo
  {pages} {190502} (\bibinfo {year} {2013})}\BibitemShut {NoStop}%
\bibitem [{\citenamefont {Ross}\ and\ \citenamefont
  {Selinger}(2014)}]{ross2014optimal}%
  \BibitemOpen
  \bibfield  {author} {\bibinfo {author} {\bibfnamefont {Neil~J.}\ \bibnamefont
  {Ross}}\ and\ \bibinfo {author} {\bibfnamefont {Peter}\ \bibnamefont
  {Selinger}},\ }\bibfield  {title} {\enquote {\bibinfo {title} {Optimal
  ancilla-free {C}lifford+{T} approximation of z-rotations},}\ }\href
  {https://arxiv.org/abs/1403.2975} {\bibfield  {journal} {\bibinfo  {journal}
  {arXiv preprint arXiv:1403.2975}} (\bibinfo {year} {2014})}\BibitemShut
  {NoStop}%
\bibitem [{\citenamefont {Bocharov}\ \emph
  {et~al.}(2015){\natexlab{a}}\citenamefont {Bocharov}, \citenamefont
  {Roetteler},\ and\ \citenamefont {Svore}}]{bocharov2015prl}%
  \BibitemOpen
  \bibfield  {author} {\bibinfo {author} {\bibfnamefont {Alex}\ \bibnamefont
  {Bocharov}}, \bibinfo {author} {\bibfnamefont {Martin}\ \bibnamefont
  {Roetteler}}, \ and\ \bibinfo {author} {\bibfnamefont {Krysta~M.}\
  \bibnamefont {Svore}},\ }\bibfield  {title} {\enquote {\bibinfo {title}
  {Efficient synthesis of universal repeat-until-success quantum circuits},}\
  }\Doi {10.1103/PhysRevLett.114.080502} {\bibfield  {journal} {\bibinfo
  {journal} {Physical Review Letters},\ }\textbf {\bibinfo {volume} {114}},\
  \bibinfo {pages} {080502} (\bibinfo {year} {2015}{\natexlab{a}})}\BibitemShut
  {NoStop}%
\bibitem [{\citenamefont {Bocharov}\ \emph
  {et~al.}(2015){\natexlab{b}}\citenamefont {Bocharov}, \citenamefont
  {Roetteler},\ and\ \citenamefont {Svore}}]{bocharov2015efficient}%
  \BibitemOpen
  \bibfield  {author} {\bibinfo {author} {\bibfnamefont {Alex}\ \bibnamefont
  {Bocharov}}, \bibinfo {author} {\bibfnamefont {Martin}\ \bibnamefont
  {Roetteler}}, \ and\ \bibinfo {author} {\bibfnamefont {Krysta~M.}\
  \bibnamefont {Svore}},\ }\bibfield  {title} {\enquote {\bibinfo {title}
  {Efficient synthesis of probabilistic quantum circuits with fallback},}\
  }\Doi {10.1103/PhysRevA.91.052317} {\bibfield  {journal} {\bibinfo  {journal}
  {Physical Review A},\ }\textbf {\bibinfo {volume} {91}},\ \bibinfo {pages}
  {052317} (\bibinfo {year} {2015}{\natexlab{b}})}\BibitemShut {NoStop}%
\bibitem [{\citenamefont {Kitaev}\ \emph {et~al.}(2002)\citenamefont {Kitaev},
  \citenamefont {Shen},\ and\ \citenamefont {Vyalyi}}]{kitaev2002classical}%
  \BibitemOpen
  \bibfield  {author} {\bibinfo {author} {\bibfnamefont {Alexei~Yu}\
  \bibnamefont {Kitaev}}, \bibinfo {author} {\bibfnamefont {Alexander}\
  \bibnamefont {Shen}}, \ and\ \bibinfo {author} {\bibfnamefont {Mikhail~N}\
  \bibnamefont {Vyalyi}},\ }\href@noop {} {\emph {\bibinfo {title} {Classical
  and quantum computation}}},\ Vol.~\bibinfo {volume} {47}\ (\bibinfo
  {publisher} {American Mathematical Society Providence},\ \bibinfo {year}
  {2002})\BibitemShut {NoStop}%
\bibitem [{\citenamefont {Marriott}\ \emph {et~al.}(2005)\citenamefont {Marriott},
\ and\ \citenamefont {Watrous}}]{MW05a}%
  \BibitemOpen
  \bibfield  {author} {\bibinfo {author} {\bibfnamefont {C.}\
  \bibnamefont {Marriott}}, \ and\ \bibinfo {author} {\bibfnamefont {J.}\
  \bibnamefont {Watrous}},\ }\bibfield  {title} {\enquote {\bibinfo {title} {
  Quantum Arthur-Merlin games},}\ } {\bibfield  {journal} {\bibinfo  {journal}
  {Computational Complexity},\ }\textbf {\bibinfo {volume} {14}},\ \bibinfo {pages}
  {122} (\bibinfo {year} {2005})}\BibitemShut {NoStop}%
\bibitem [{\citenamefont {Wecker}\ \emph {et~al.}(2014)\citenamefont {Wecker},
  \citenamefont {Bauer}, \citenamefont {Clark}, \citenamefont {Hastings},\ and\
  \citenamefont {Troyer}}]{WBCH13a}%
  \BibitemOpen
  \bibfield  {author} {\bibinfo {author} {\bibfnamefont {Dave}\ \bibnamefont
  {Wecker}}, \bibinfo {author} {\bibfnamefont {Bela}\ \bibnamefont {Bauer}},
  \bibinfo {author} {\bibfnamefont {Bryan~K.}\ \bibnamefont {Clark}}, \bibinfo
  {author} {\bibfnamefont {Matthew~B.}\ \bibnamefont {Hastings}}, \ and\
  \bibinfo {author} {\bibfnamefont {Matthias}\ \bibnamefont {Troyer}},\
  }\bibfield  {title} {\enquote {\bibinfo {title} {Gate-count estimates for
  performing quantum chemistry on small quantum computers},}\ }\Doi
  {10.1103/PhysRevA.90.022305} {\bibfield  {journal} {\bibinfo  {journal}
  {Physical Review A},\ }\textbf {\bibinfo {volume} {90}},\ \bibinfo {pages}
  {022305} (\bibinfo {year} {2014})}\BibitemShut {NoStop}%
\bibitem [{\citenamefont {Poulin}\ \emph {et~al.}(2015)\citenamefont {Poulin},
  \citenamefont {Hastings}, \citenamefont {Wecker}, \citenamefont {Wiebe},
  \citenamefont {Doherty},\ and\ \citenamefont {Troyer}}]{poulin2015trotter}%
  \BibitemOpen
  \bibfield  {author} {\bibinfo {author} {\bibfnamefont {David}\ \bibnamefont
  {Poulin}}, \bibinfo {author} {\bibfnamefont {Matthew~B.}\ \bibnamefont
  {Hastings}}, \bibinfo {author} {\bibfnamefont {Dave}\ \bibnamefont {Wecker}},
  \bibinfo {author} {\bibfnamefont {Nathan}\ \bibnamefont {Wiebe}}, \bibinfo
  {author} {\bibfnamefont {Andrew~C.}\ \bibnamefont {Doherty}}, \ and\ \bibinfo
  {author} {\bibfnamefont {Matthias}\ \bibnamefont {Troyer}},\ }\bibfield
  {title} {\enquote {\bibinfo {title} {The trotter step size required for
  accurate quantum simulation of quantum chemistry},}\ }\href
  {http://www.rintonpress.com/xxqic15/qic-15-56/0361-0384.pdf} {\bibfield
  {journal} {\bibinfo  {journal} {Quantum Information and Computation},\
  }\textbf {\bibinfo {volume} {15}},\ \bibinfo {pages} {361--384} (\bibinfo
  {year} {2015})}\BibitemShut {NoStop}%
\bibitem [{\citenamefont {Low}\ and\ \citenamefont
  {Chuang}(2017){\natexlab{b}}}]{low2017usa}%
  \BibitemOpen
  \bibfield  {author} {\bibinfo {author} {\bibfnamefont {Guang~Hao}\
  \bibnamefont {Low}}\ and\ \bibinfo {author} {\bibfnamefont {Isaac~L.}\
  \bibnamefont {Chuang}},\ }\bibfield  {title} {\enquote {\bibinfo {title}
  {Hamiltonian simulation by uniform spectral amplification},}\ }\href
  {https://arxiv.org/abs/1707.05391} {\bibfield  {journal} {\bibinfo  {journal}
  {arXiv preprint arXiv:1707.05391}} (\bibinfo {year}
  {2017}{\natexlab{b}})}\BibitemShut {NoStop}%
\bibitem [{\citenamefont {Harrow}\ \emph {et~al.}(2009)\citenamefont {Harrow},
  \citenamefont {Hassidim},\ and\ \citenamefont {Lloyd}}]{HHL09a}%
  \BibitemOpen
  \bibfield  {author} {\bibinfo {author} {\bibfnamefont {Aram~W.}\ \bibnamefont
  {Harrow}}, \bibinfo {author} {\bibfnamefont {Avinatan}\ \bibnamefont
  {Hassidim}}, \ and\ \bibinfo {author} {\bibfnamefont {Seth}\ \bibnamefont
  {Lloyd}},\ }\bibfield  {title} {\enquote {\bibinfo {title} {Quantum algorithm
  for solving linear systems of equations},}\ }\Doi{10.1103/PhysRevLett.103.150502}\href@noop {} {\bibfield
  {journal} {\bibinfo  {journal} {Phys. Rev. Lett.},\ }\textbf {\bibinfo
  {volume} {103}},\ \bibinfo {pages} {150502} (\bibinfo {year}
  {2009})}\BibitemShut {NoStop}%
\end{thebibliography}

%

\end{document}